\documentclass[9pt,twocolumn,twoside]{osajnl}

\usepackage{mathtools}

\DeclareMathOperator{\erfc}{erfc}
\DeclareMathOperator{\erf}{erf}
\journal{ao} 
\setboolean{shortarticle}{false}

\title{Probabilistic model for spatially acquiring optical links in space under influence of band-limited beam jitter }
\author[1]{Gerald Hechenblaikner}
\affil[1]{Airbus Defence and Space, Claude-Dornier-Strasse, 88090 Immenstaad, Germany}

\begin{abstract}
An analytical model is derived for the probability of failure (P-fail) to spatially acquire an optical link with a jittering search beam.
The analytical model accounts for an arbitrary jitter spectrum and considers the associated correlations between jitter excursions on adjacent tracks of the search spiral.
An expression of P-fail in terms of basic transcendental functions is found by linearizing the exact analytical model with respect to the correlation strength.
Predictions from the models indicate a strong decrease of P-fail with increasing correlation-strength, which is found to be in excellent agreement to results from Monte Carlo simulations. The dependency of P-fail on track-width and scan speed is investigated, confirming previous assumptions on the impact of correlations. Expressions and applicable constraints are derived for the limits of full and no correlations, and the optimal track width to minimize the acquisition time is computed for a range of scan speeds.
The model is applicable to optical terminals equipped with a fast beam steering mirror, as often found for optical communication missions in space.
\end{abstract}

\setboolean{displaycopyright}{false}

\begin{document}

\maketitle

\section{Introduction}
\label{sec:intro}
Optical communication in space and from space to ground is becoming increasingly important in order to fulfill requirements on larger band-width and global availability for vast amounts of data \cite{hauschildt2019global}.
While radio-frequency (RF) links have so far been able to deliver most of the communication needs, they will not be able to keep up with the growing demand. As optical links offer the prospect of much higher channel capacity and transmission speed compared to RF links, they are likely to succeed or at least complement traditional RF communication systems in many applications \cite{kaushal2016optical}.
Numerous missions during the past two decades spear-headed the initial technology developments and paved the way for increasingly robust link deployments (see e.g. \cite{fletcher1991silex, jono2006demonstrations, fields2009nfire,benzi2016optical,hauschildt2019global}, and an overview in \cite{hemmati2020near,kaushal2016optical}).\\
As an added benefit, optical links are more secure \cite{hauschildt2019global} because the transmitted beams have narrow divergence angles and are targeted to a comparatively small region so that they cannot be easily intercepted. However, this benefit turns out to be a major disadvantage when it comes to establishing the link in the first place. The small beam-width makes it difficult to ``hit'' the receiver with the transmitted beam, considering that the uncertainties in the beam pointing are typically 1-2 orders of magnitude larger than the beam width, unless a dedicated wide field search beacon is used. For this reason, the ``initial beam acquisition'' is one of the most critical challenges for optical communication and a robust acquisition the pre-requisite for successful deployment of optical links.

\subsection{Link Acquisition Process}
\begin{figure}
\centerline{\includegraphics[width=\columnwidth]{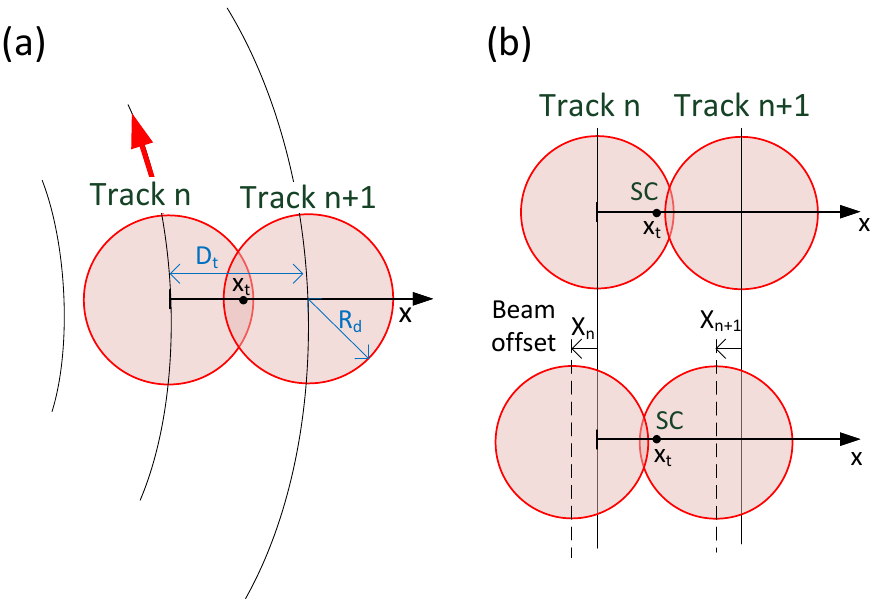}}
\caption  {(a) The scanning beam moves along the spiral track (excerpt shown) and has a ``capture region'' defined by the red-shaded area within a distance $R_d$ from the beam center. There is an overlap between the beams on two adjacent tracks (dark shaded region). (b) Top: In absence of jitter, the SC is within the capture region on track n.  Bottom: In presence of jitter, the beam is displaced by $X_n$ and not detected on track n. However, the beam is detected on track n+1, provided the displacement $X_{n+1}$ remains constant  (correlated). \label{fig_1}}
\end{figure}

This section gives a short description of a typical acquisition process as background information (see e.g. \cite{hemmati2020near} for details).
Owing to the large uncertainty of the position of the receiving spacecraft (SC) and the direction of the optical transmission axis, the transmitting SC1 performs a search scan of the uncertainty region by guiding the transmitted beam along an Archimedean spiral.  The receiving SC2, indicated by the small black circle of Fig.~\ref{fig_1}(a), is located somewhere in this region in between two spiral tracks which are separated by the track width $D_t$. SC2 is assumed to detect the beam if its distance to the beam center is smaller than the radius of detection $R_d$. In order to guarantee good coverage of the scanning area, the track width $D_t$ is chosen such that there is an overlap $OV=2R_d-D_t$ between the beam footprints (red-shaded area) on adjacent tracks which must be sufficiently large to account for deviations from the nominal path due to beam jitter.\\
When the scanning beam sweeps across the location of SC2 in the uncertainty plane, SC2 is assumed to detect it with sufficient accuracy (e.g. on a matrix detector) to re-orient its transmitted beam in the direction of the received light. As soon as SC1, which has meanwhile continued scanning the uncertainty region, receives the transmitted beam from SC2, it stops scanning and likewise re-directs its transmitted beam into the direction of the received light from SC2.
Once both SC have spatially acquired the received beams of their counterpart, one SC may lock its laser to the frequency of the received light through a phase-locked loop in order to facilitate coherent data transmission. 

\subsection{Overview of models and assumptions}
\label{sec:analytical_models_overview}
For optical link acquisition, the scanning beam of SC 1 is typically moved along an Archimedean search spiral with constant angular velocity, covering the uncertainty distribution of the receiving SC2 from the center to the maximum scan radius (see e.g.\cite{benzi2016optical, friederichs2016vibration, ma2021satellite, sterr2011beaconless, li2011analytical, hu2022multi}). Considering the Gaussian profile of the uncertainty distribution (see section 2\ref{sec:uncertainty_distribution}), the regions of higher probability are scanned before those of lower probability, making the Archimedean spiral a very efficient choice. A detailed comparison between spiral, raster, Lissajous, and Rose search patterns is given in \cite{scheinfeild2001}.

\begin{figure}
\centerline{\includegraphics[width=\columnwidth]{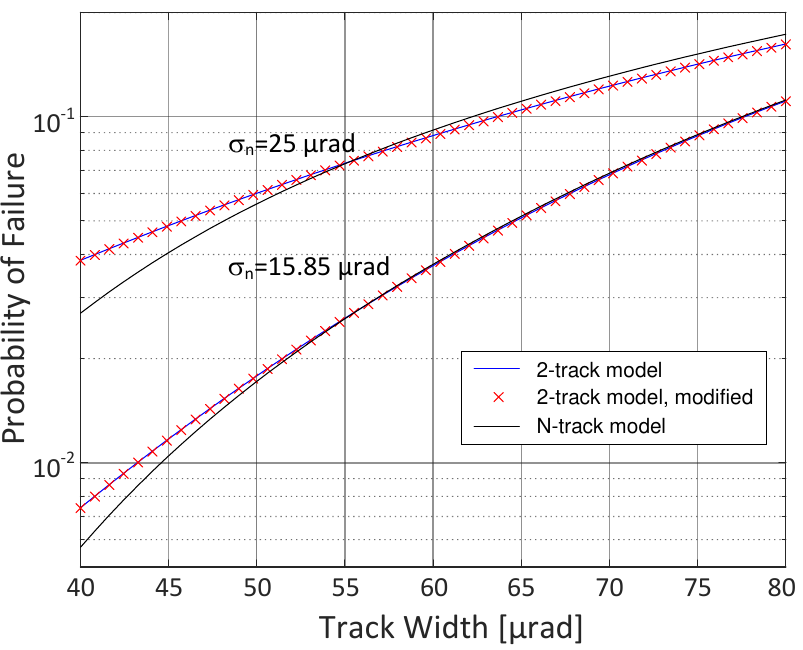}}
\caption  {Probability of failure plotted against track-width for typical mission parameters used in this paper (see Tab.~\ref{tab:simulation_parameters}), with RMS beam jitter either $15.85\,\mu rad$ (lower curves) or $25\,\mu rad$ (upper curves). Analytical predictions given for the ``2-track model'' \cite{hechenblaikner2021analysis} (blue line), the modified 2-track model \cite{hu2022multi} (red crosses), and the ``N-track model'' \cite{hechenblaikner2022impact} for $N=7$ (black line).\label{fig_2}}
\end{figure}

Previous analytical models to compute the probability of acquisition failure ($P_{fail}$) ignored correlations of the beam jitter, although this is a dominating effect for fast scanning terminals, which motivates the development of an entirely new analytical model in this paper. Below, important aspects that characterize acquisition models are summarized:

\begin{enumerate}
\item {\textbf{Neglecting tangential jitter:} In order to simplify the derivation of an analytical model, it is often sufficient to account for radial jitter only (across the spiral track). Neglecting the tangential degree-of-freedom (dof) is valid as long as the total jitter root-mean-square (RMS) does not exceed the beam diameter \cite{hechenblaikner2022impact}. 
The authors of \cite{friederichs2016vibration} included the tangential dof in the derivation of an analytical model, but this comes at the expense of much more sophisticated mathematics with little benefit in terms of accuracy. In fact, it has been shown that the results of Monte Carlo simulations yield almost identical results if radial jitter alone is applied compared to the case where radial and tangential jitter are applied\cite{hechenblaikner2022impact,hechenblaikner2023optical}. These simulations also were in close agreement to predictions of a 1-dimensional (1d) analytical model \cite{hechenblaikner2021analysis} that only accounts for radial jitter, which further verified the validity of neglecting tangential jitter.}

\item{\textbf{Model dimensionality:} Using a 1d analytical model for the acquisition probability of failure $P_{fail}$ also amounts to assuming that the radius of curvature of the tracks does not enter the computation of $P_{fail}$. This further implies that the computed failure probability for a specific SC location in between two tracks n and n+1 does not depend on the value of n 
but only on the radial offset $\Delta_r$ from track n, which is exactly what was observed in \cite{friederichs2016vibration} and proven by the close match of 2d Monte Carlo simulations to the predictions of a 1d analytical model in \cite{hechenblaikner2022impact}.}

\item{\textbf{Uncertainty distribution between tracks:} As the uncertainty distribution of where to find the SC in the scanning plane is generally changing only slowly on the scale of the track-width $D_t$, it can be approximated to be constant and given by its mean value in between any two tracks n and n+1. In fact, it can be shown that probabilities calculated with this approximation are exact for changes up to linear order of the uncertainty distribution. 
Therefore, the total probability of failure $P_{fail}$ can be found by integrating the conditional probability of missing the scanning beam for a given offset $x_t$ from track n over a uniform density in between the two tracks \cite{hechenblaikner2021analysis}.
A recent paper \cite{hu2022multi} suggested to improve the accuracy of the calculation by piecewise integration of the conditional probability over all track pairs, density weighted by the exact shape of the uncertainty distribution in each segment. However, as can be seen in Fig.~\ref{fig_2}, there is no noticeable difference between the results of the much more involved approach of \cite{hu2022multi}(red crosses) and the original model of \cite{hechenblaikner2021analysis} (blue line). The latter model yields accurate results, if the width of the uncertainty distribution $\sigma_{uc}$ is larger than $D_t$, which is the  case for all conceivable mission scenarios.}

\item{\textbf{Detection on multiple tracks:} Previous models \cite{friederichs2016vibration,ma2021satellite} have not accounted for the fact that if the SC fails to detect the jittering beam on the scanning track closest to it, there is still a reasonable chance it may detect it on one of the adjacent tracks. In fact, accounting for this possibility is extremely important as this effect typically dominates the predicted probability of failure.
This effect was first included in a model where detection may occur on either track n or track n+1 (``2-track model'') which is accurate for small jitter and if the track width $D_t$ is larger than the radius of detection $R_d$ \cite{hechenblaikner2021analysis}. For larger jitters and/or narrower track widths, the model must be extended to account for potential detection of up to $N$ neighboring tracks (``N-track model'') \cite{hechenblaikner2022impact}. This is also demonstrated in Fig.~\ref{fig_2} which shows that the probability of failure for the ``N-track model'' (black curve) deviates from the one for the ``2-track model'' (blue curve)  at small track widths, and the deviations are greater for larger RMS jitter ($\sigma_n=25.00\,\mu rad$) than for small jitter ($\sigma_n=15.85\,\mu rad$).}

\item{\textbf{Beam detection model:} 
Previous analyses \cite{friederichs2016vibration,ma2021satellite,hechenblaikner2021analysis} assumed that the scanning beam is detected by the SC if the distance between the beam center and the SC is smaller than the radius of detection $R_d$ at any any time during the spiral scan. This detection model is referred to as the `Hard-Sphere Model'' (HSM), due to its simple geometric interpretation (see also Fig.~\ref{fig_1}). It is remarkable that the HSM, which combines all detection characteristics in a single parameter ($R_d$), yields very similar results to those of (much more sophisticated) physically representative models which simulate the integration across the actual beam intensity distribution during a finite integration time and account for detector noise sources and a detection threshold corresponding to a specific signal-to-noise-ratio (SNR) \cite{hechenblaikner2022impact, hechenblaikner2023optical}. Unfortunately, this agreement breaks-down if RMS amplitude and band-width of the beam jitter are very large \cite{hechenblaikner2022impact}. However, the HSM is fully applicable to the parameter regime covered in this paper.}

\item{\textbf{Jitter correlation effects:} The ``N-track model'' yields accurate predictions in agreement to those of Monte Carlo simulations as long as certain assumptions on the jitter correlation are met. These assumptions are (1) that the jitter is correlated on a time-scale of the beam passage time $T_{pass}$ during which the scanning beam sweeps over the SC location and (2) that the jitter between two adjacent tracks is uncorrelated, i.e. the  jitter is uncorrelated on the time-scale  of one spiral revolution $T_{rev}$, which may be expressed as an inequality for the width of the jitter auto-correlation function $\tau_0$ as follows: $T_{pass}<\tau_0<T_{rev}$. As discussed in \cite{hechenblaikner2023optical}, the actual failure rate $P_{fail}$ drops below the one predicted by the ``N-track'' model for $\tau_0<T_{pass}$ (as integration windows become uncorrelated) and for $\tau_0>T_{rev}$ (as jitter between tracks becomes correlated).}
\end{enumerate}

It is interesting to compare the regime of validity between the ``N-track model'' and the analytical model of \cite{friederichs2016vibration} which the authors specifically applied to the European Data Relay System (EDRS). In their model, the authors assumed white Gaussian noise of unlimited bandwidth for the beam jitter which has the following consequences: (1) Unlimited band-width implies infinitely small correlation times so that the primary effect discussed in this paper, the impact of jitter correlations between tracks, cannot emerge, no matter how fast the beam is scanning. (2) Infinitely small correlation times also imply that the jitter fluctuations which occur within one integration window become uncorrelated with the jitter fluctuations of all other integration windows during $T_{pass}$. This leads to much lower compound failure probabilities than for the ``N-track model'', where fluctuations are assumed correlated between the windows and jitter displacements are almost equal for all windows during $T_{pass}$. The model in \cite{friederichs2016vibration} therefore applies under the condition that $\tau_0\ll T_{pass}<T_{rev}$. 
However, for an optical communication mission with a comparatively stable platform, such as the one of SILEX \cite{hemmati2020near}, and an optical terminal scanning close to its maximum speed \cite{sterr2011beaconless}, the rather opposite regime applies.

\subsection{Scope and structure of this paper}
In section \ref{sec:corr_analytical_model} an analytical model is derived to address this regime of fast scanning optical terminals, where $T_{pass}< T_{rev}\leq\tau_0$ and jitter fluctuations are strongly correlated between adjacent spiral tracks. 
For simplicity, the model is restricted to the case where detection may occur either on track 'n' or track 'n+1' (``2-track model''), thus requiring that the RMS jitter is smaller than the radius of detection and the track-width ($\sigma_n < R_d, D_t$).
Using this model, the impact of increasing correlations, e.g. due to increasing scan speed, on the probability of failure $P_{fail}$ is accurately calculated. Discussing the model predictions for realistic parameters in section 3\ref{sec:Variation_gamma}, it becomes apparent that for fast scan speeds the emerging correlations may reduce $P_{fail}$ by orders of magnitude, which is a dominating effect that is not accounted for by previous analytical models. Remarkably good agreement is found between the analytical model predictions and Monte Carlo simulations which confirm the model validity and the interpretation with respect to correlation effects. In section 3\ref{sec:parametric_relationships} it is demonstrated how the value of a single compound parameter (which only depends on scan speed, bandwidth of the jitter spectrum, and width of the uncertainty distribution) fully determines the impact of correlations which is compared to the uncorrelated case using a dedicated performance metric. Finally, the optimal spiral track-width and associated beam overlap which minimize the mean acquisition search time are determined for a given scan speed in section 3\ref{sec:mean_acquisition_time}, accounting for the possibility that a scan is repeated in case of failure.

\section{The ``correlated analytical model''}
\label{sec:corr_analytical_model}
The aim of this section is to derive an analytical model for the probability of acquisition failure due to beam jitter which accounts for correlations of the jitter between adjacent tracks and which is henceforth referred to as ``correlated analytical model''.
To this end, the ``2-track model'' of \cite{hechenblaikner2021analysis} is extended to include correlation effects. 
Where appropriate, similar definitions of physical parameters, coordinate systems, and observables, are used, referring the reader to \cite{hechenblaikner2021analysis} for more detailed information.
Instead, the focus in this section lies on the distinctive assumptions and new mathematical concepts required for the derivation of the ``correlated analytical model''.
At first, some basic physical quantities and baseline parameters that are needed for the actual derivation in subsection 2\ref{sec:model_derivation} are introduced.

\subsection{The SC uncertainty distribution}
\label{sec:uncertainty_distribution}
For successful acquisition, SC1 must be accurately pointed towards the receiving SC2, considering that the direction of the transmitted beam should be slightly pointed ahead from the direct line of sight between the two SC (``point-ahead-angle'') due to the finite light propagation time. In practice, a number of errors affect the pointing, among which the primary groups are knowledge errors of the transceiver optical axis with respect to the SC reference frame (caused by calibration error and ground-to-orbit shifts), ephemeris and orbit determination errors, and SC attitude knowledge errors. Certain instabilities, such as thermo-elastic drifts, may affect both,  the alignment of the SC attitude sensor (impacting the SC attitude knowledge error), and the alignment of the transceiver optical axis (impacting the optical axis error). 
All errors are typically combined by adding their respective variances, assuming that the underlying effects are uncorrelated, to find the total variance $\sigma_{uc}^2$ which is assumed to be identical for azimuth and elevation angles. The uncertainty distribution is then found as the product of two normal distributions (one for each angle), which can be represented by a Rayleigh distribution along the radial direction of the uncertainty plane \cite{scheinfeild2000}:
\begin{equation}
PDF_{uc}(r)=\frac{r}{\sigma_{uc}^2}e^{-\frac{r^2}{2\sigma_{uc}^2}},
\label{eq:Rayleigh}
\end{equation}

\subsection{Platform jitter spectrum}
During the acquisition the satellite platform vibrates and leads to jitter with a variance of $\sigma_n^2$. For the following analysis, the one-sided Power Spectral Density (PSD) of the jitter spectrum specified by ESA for the SILEX platform is assumed that is given by \cite{hemmati2020near, teng2018optimization}
\begin{equation}
S(f)=\frac{160~\mu\mathrm{rad^2 Hz^{-1}}}{1+(f/f_{r})^2},\label{eq:jitter_spectrum}
\end{equation}
where the $f_r=1\,\mathrm{Hz}$ is the roll-off frequency beyond which the spectral density drops off as $1/f^2$. The RMS fluctuations and jitter variance $\sigma_n^2$ are found from the power spectral density to be:
\begin{equation}
RMS_{jitter}=\sqrt{\sigma_n^2}=\sqrt{\int_0^{\infty}df S(f)}=15.85\,\mu rad
\end{equation}

\subsection{Acquisition Parameter Overview}
In order to obtain predictions from the analytical model and compare them to the results of Monte Carlo simulations, the physical configuration parameters given in Table \ref{tab:simulation_parameters} are used.
The jitter spectrum is based on the one specified for the SILEX mission in Eq.\,\ref{eq:jitter_spectrum}, the maximum scan speed was extracted from data for optical terminals given in \cite{sterr2011beaconless, friederichs2016vibration}, and the radius of detection $R_d=40\,\mu rad$ is close to the effective beam divergence (half) angle of a small telescope of 2.8 cm aperture diameter, where the transmitted beam is clipped at a radius of 1.5 times the beam waist. Note that the maximum possible scan speed is typically limited by the performance of the fine-steering mirror with associated control electronics.
\begin{table}[h]
\centering
\caption{\label{tab:simulation_parameters}
Default mission parameters used for simulations}
\begin{tabular}{lll}
\hline
Symbol & Name & Value\\
\hline
$R_d$ & radius of detection & $40\,\mu rad$\\
$D_t$ & track width & $62.8\,\mu rad$ \\
$\sigma_n$ & RMS beam jitter (per dof) & $15.85\,\mu rad$\\
$f_r$ & spectral roll-off frequency & $1\,Hz$ \\
$\sigma_{uc}$ & uncertainty distr. width & $290.7\,\mu rad$\\
$R_{uc}$ & max. scan radius & $1000\,\mu rad$ \\
$\gamma_{max}$ & max. scan speed & $70\,mrad/s$ \\
$\gamma_{min}$ & min. scan speed for correlation effects & $14\,mrad/s$\\
\end{tabular}
\end{table}

\subsection{Model Derivation}
\label{sec:model_derivation}
In this section an analytical expression is derived for the acquisition probability of failure $P_{fail}$ in the presence of beam jitter that is correlated in between tracks. This derivation generalizes the one presented in \cite{hechenblaikner2021analysis} by accounting for the effect of correlations, which allows to accurately predict the dependency of the probability of failure $P_{fail}$ on the scan speed $\gamma$ and and normalized jitter auto-correlation function $r_{xx}(\tau)$. Qualitative arguments for why such a dependency exists were previously provided in \cite{hechenblaikner2022impact}. These arguments were motivated to some extent by analyzing the results of Monte Carlo simulations, but the assumptions could not be directly verified. In contrast, in this paper the direct parametric relationship is obtained between the acquisition failure rate and the physical system parameters through the derivation of an ab initio model whose predictions are found to be in excellent agreement to those of Monte Carlo simulations.\\
For the derivation it is assumed that the SC is located at a random position $X_t$ in between two tracks n and n+1 of the search spiral, where the coordinate system $x$ starts at track n and radially points towards track n+1 (see Fig.~\ref{fig_1}a). As discussed in section 1\ref{sec:analytical_models_overview}, the SC uncertainty distribution only changes slightly and can therefore assumed to be constant in between the two tracks, if $\sigma_{uc}> D_{t}$. The probability density function $u(x_t)$ is then given by:
\begin{equation}
u(x_t)=\frac{2}{D_t}\hspace{1cm} 0\leq x_t \leq D_t/2\label{eq:constant_density}
\end{equation}
Note that because of the underlying symmetry of this model it is only necessary to consider potential locations of the SC in between $[0, D_t/2]$ instead of $[0,D_t]$, which slightly reduces the computational effort. As discussed in section 1\ref{sec:analytical_models_overview}, the error made by assuming a constant density in this small interval is completely negligible for a typical mission scenario.\\
The jittering beam can only be detected on either track n or track n+1 with reasonable probability, provided that the jitter is sufficiently small and the track width sufficiently large compared  to the beam radius of detection $R_d$. If the jitter becomes larger and/or the track width becomes narrower, there is a chance that the large jitter excursions lead to a detection on tracks with a larger radial distance (e.g. tracks n-1 or n+2), in which case the ``2-track model'', originally introduced in \cite{hechenblaikner2021analysis}, must be extended to ``N-tracks'', as shown in \cite{hechenblaikner2022impact}.\\
The event that the scanning beam is not detected on track n shall be referred to as event A and that it is not detected on track n+1 as event B, respectively.
Quite generally, the beam is said to be detected if the distance between the beam center and the location of the SC is smaller than the radius of detection $R_d$.
Jitter fluctuations that occur while the beam is passing by the SC on track n are modeled by a random variable $X_n$ which follows a normal distribution of zero mean and variance $\sigma_n^2$: $X_n \sim\mathcal{N}(0,\sigma_n^2)$. The associated probability density function (PDF) is given by
\begin{equation}
g(x_n)=\frac{1}{\sqrt{2\pi}\sigma_n} e^{-\frac{x_n^2}{2\sigma_n^2}}\label{eq:normal_distribution}
\end{equation}
Similarly, fluctuations that occur after one spiral revolution, while the beam is passing by the SC on the adjacent track n+1, are modeled by an identically distributed random variable $X_{n+1} \sim\mathcal{N}(0,\sigma_n^2)$. Based on the above definitions, the following probabilistic equations for events A and B can be derived \cite{hechenblaikner2021analysis}:
\begin{eqnarray}
P(A)&=&P(X_n>R_d-X_t)\nonumber\\
P(B)&=&P(X_{n+1}<D_t-R_d-X_t).\label{eq:prob_inequality}
\end{eqnarray}
The total probability $P(A\,and\,B)$ that the scanning beam is neither detected on track n nor on track n+1 is derived considering that Eqs.~\ref{eq:prob_inequality} are correlated through the SC position $X_t$ as well as the jitter fluctuations $X_n, X_{n+1}$ for which the general assumption is that $\langle X_n X_{n+1}\rangle \neq 0$. Note that jitter fluctuations are generally correlated, if the width $\tau_0$ of the normalized jitter auto-correlation function $r_{xx}(\tau)$ is larger than the mean spiral revolution time $T_{mean}$:
\begin{equation}
\tau_0> T_{mean}.\label{eq:correlation_condition}
\end{equation}
From the spectrum given in Eq.\ref{eq:jitter_spectrum} the normalized auto-correlation function is found to decay exponentially and its $1/e$ width $\tau_0$ is determined\cite{hechenblaikner2022impact}:
\begin{align}
r_{xx}(\tau)&=e^{-\frac{|\tau|}{\tau_0}},\label{eq:auto_correlation}\\
\tau_0&:=\frac{1}{2\pi f_r}.\label{eq:ac_width}
\end{align}
The mean radial distance of the SC is found from the uncertainty distribution of Eq.~\ref{eq:Rayleigh} to be $R_{mean}=\sqrt{\pi/2}\sigma_{uc}$ from which the mean spiral revolution time is obtained:
\begin{equation}
T_{mean}=\frac{2\pi R_{mean}}{\gamma}=\sqrt{\frac{\pi}{2}}\frac{2\pi\sigma_{uc}}{\gamma}.\label{eq:T_mean}
\end{equation}
Inserting Eqs.~\ref{eq:T_mean} and \ref{eq:ac_width} into Eq.~\ref{eq:correlation_condition}, an expression for the minimum scan speed $\gamma_{min}$ is obtained where correlation effects between tracks n and n+1 should emerge \cite{hechenblaikner2022impact}
\begin{equation}
\gamma_{min}=f_r\sqrt{\pi/2}\left(2\pi\right)^2\sigma_{uc},
\end{equation}
which is evaluated to be $\gamma_{min}=14\,mrad/s$ for the default mission parameters of Table~\ref{tab:simulation_parameters}.
Considering the correlations between Eqs.~\ref{eq:prob_inequality}, the conditional probability is derived that event A occurs given that the SC is at a certain location $X_t=x_t$ and the jitter fluctuation on track n has a size $X_n=x_n$:
\begin{equation}
P(A|X_t=x_t, X_n=x_n)=H\left(x_n-(R_d-x_t)\right),\label{eq:cond_corr_PA}
\end{equation}
where $H(x)$ is the Heaviside function.
Integrating the conditional probability over the density functions for $X_n$ and $X_t$ the probability $P(A)$ is obtained that the SC is not detected on track n:
\begin{eqnarray}
P(A)&=&\int_{-\infty}^{\infty}d x_n g(x_n)\int_{0}^{D_t/2} dx_t\frac{2}{D_t} H\left(x_n-(R_d-x_t)\right)\nonumber\\
&=& \frac{2}{D_t} \int_{0}^{D_t/2} dx_t\frac{1}{2}\erfc\left(\frac{R_d-x_t}{\sqrt{2}\sigma_n}\right)
\end{eqnarray}
In order to calculate the probability $P(B)$ that the SC is not detected on track n+1 it is now considered that the jitter on track n+1, described  by the random variable $X_{n+1}$, is correlated with the jitter $X_n$ on track n. Therefore, $X_{n+1}$ is decomposed into a contribution of amplitude $\delta_r$ that is fully correlated with $X_n$ and a contribution $X_{n,\perp}$ of amplitude $\epsilon_r=(1-\delta_r^2)^{1/2}$ that is orthogonal to $X_n$ and also follows a normal distribution $X_{n,\perp} \sim\mathcal{N}(0,\sigma_n^2)$. To this end, the correlation amplitude $\delta_r$ is found from the normalized auto-correlation function of Eq.~\ref{eq:auto_correlation} that must be evaluated for the spiral revolution time on track n which is assumed to have a radius $r$: 
\begin{align}
\delta_r&=r_{xx}(2\pi r/\gamma)=e^{-\beta r}\label{eq:delta_r}\\
\beta&:= \frac{\left(2\pi\right)^2f_r}{\gamma}.\label{eq:beta}
\end{align}
This yields:
\begin{equation}
X_{n+1}=\delta_r X_n+\epsilon_r X_{n,\perp}.\label{eq:delta_decomposition}
\end{equation}
The conditional probability $P(B)$ of not detecting the scanning beam on track n+1, given that the SC is at a certain location $X_t=x_t$ and the jitter fluctuation on track n has a size $X_n=x_n$, is found to be:
\begin{align}
& P(B|X_t=x_t, X_n=x_n)=P(\delta_r x_n+\epsilon_r X_{n,\perp}<D_t-R_d-x_t)\nonumber\\
&=P(X_{n,\perp}<\frac{1}{\epsilon_r}\left(D_t-R_d-x_t-\delta_r x_n\right))\nonumber\\
&=\int^{\frac{1}{\epsilon_r}\left(D_t-R_d-x_t-\delta_r x_n\right)}_{-\infty} dx_{n,\perp}g(x_{n,\perp})\nonumber\\
&=\frac{1}{2}\erfc\left(\frac{R_d-D_t+x_t+\delta_r x_n}{\epsilon_r\sqrt{2}\sigma_n}\right)\label{eq:cond_corr_PB}
\end{align}
The total probability $P(A\,and\,B)$ is then found by integrating the product of the conditional probabilities of Eqs.\,\ref{eq:cond_corr_PA},\ref{eq:cond_corr_PB} over the marginal variables $x_t$ and $x_n$ which must be weighted by their respective density functions $u(x_t)$ and $g(x_n)$, given in Eqs.~\ref{eq:constant_density} and \ref{eq:normal_distribution}, respectively:
\begin {align}
 &P_{fail}(\delta_r)=P(A\,and\,B)\nonumber\\
&=\int_0^{D_t/2}dx_t\frac{2}{D_t}\int_{-\infty}^{\infty} dx_n \left[g(x_n) P(A|X_t=x_t, X_n=x_n)\right.\nonumber\\
&\hspace{1cm}\left.\times P(B|X_t=x_t, X_n=x_n)\right]\nonumber\\
&=\frac{2}{D_t}\int_0^{D_t/2}dx_t\int_{R_d-x_t}^{\infty} dx_n g(x_n)\frac{1}{2}\erfc\left(\frac{R_d-D_t+x_t+\delta_r x_n}{\epsilon_r\sqrt{2}\sigma_n}\right),\label{eq:total_corr_prob}
\end{align}
where the dependency of the failure rate $P_{fail}(\delta_r)$ on the correlation amplitude $\delta_r$ of Eq.~\ref{eq:delta_r} is explicitly noted. The latter depends itself on the spiral radius $r$ of track n.
However, as the SC could be located at any radius $r$, the average of Eq.~\ref{eq:total_corr_prob} must be taken over the SC uncertainty distribution of Eq. \ref{eq:Rayleigh} to obtain the average probability of failure:
\begin{equation}
P_{fail,av}=\int_0^{\infty} dr P_{fail}\left(\delta_r\right)\frac{r}{\sigma_{uc}^2}e^{-\frac{r^2}{2\sigma_{uc}^2}}
\label{eq:average_P_fail}
\end{equation}
As mentioned in section 1\ref{sec:analytical_models_overview}, there is no improvement in accuracy to be gained if the density function of the SC location $x_t$ is not assumed to be constant but given by the exact Rayleigh distribution when performing the integral over $x_t$ in Eq.~\ref{eq:total_corr_prob}, as the Rayleigh distribution is only slightly changing over the short distance between tracks and linear changes are averaged out. However, the correlation amplitude $\delta_r$ for a given spiral radius greatly reduces towards larger radii, which in turn leads to strongly increasing failure rates. Therefore, it is very important to average the dependency of the failure rate on $\delta_r$ across the uncertainty distribution as done in Eq.~\ref{eq:average_P_fail}.\\
It is also possible to calculate the mean correlation amplitude $\delta_{mean}$ which is defined as the average of $\delta_r$ over the uncertainty distribution. Using $\delta_{mean}$ to evaluate $P_{fail}(\delta_{mean})$ through Eq.~\ref{eq:total_corr_prob}  yields probabilities of failure that are quite close to those obtained from Eq.~\ref{eq:average_P_fail}.

\subsection{The extreme limits of full and no correlations}
It is interesting to investigate the extreme limit that $X_n$ is fully correlated with $X_{n+1}$. In this case it is assumed that $X_{n+1}=X_n$ and the following condition for acquisition failure is found from Eqs.\,\ref{eq:prob_inequality}:
\begin{equation}
D_t-R_d-X_t>X_n>R_d-X_t\hspace{0.5cm}\rightarrow\hspace{0.5cm} D_t>2R_d.\\
\end{equation}
This means that acquisition only fails if there is no overlap between beams on adjacent tracks, i.e. if $OV=2R_d-D_t<0$. If, on the contrary, there is an overlap and $2R_d-D_t>0$, it follows that acquisition must always succeed and $P_{fail}=0$, irrespective of all other physical parameters such as the jitter power. This result can be understood when considering that fully correlated jitter on adjacent tracks amounts to a bias offset, i.e. a constant displacement of the beam from the nominal track, which has no impact on the probability of failure, as shown in Fig.~\ref{fig_1}(b). It is only bias drifts which lead to failure \cite{hechenblaikner2022impact}.\\
The other extreme of uncorrelated jitter between tracks implies that $\langle X_n X_{n+1}\rangle=0$ and therefore $\delta_r=0$ and $\epsilon_r=1$.
Under these assumptions Eq.~\ref{eq:total_corr_prob} yields:
\begin{align}
P_{fail,uc}&=\frac{2}{D_t}\int_0^{D_t/2}dx_t\int_{R_d-x_t}^{\infty} dx_n g(x_n)\frac{1}{2}\erfc\left(\frac{R_d-D_t+x_t}{\sqrt{2}\sigma_n}\right)\nonumber\\
&=\frac{2}{D_t}\int_0^{D_t/2}dx_t\frac{1}{2}\erfc\left(\frac{R_d-x_t}{\sqrt{2}\sigma_n}\right)\frac{1}{2}\erfc\left(\frac{R_d-D_t+x_t}{\sqrt{2}\sigma_n}\right),\label{eq:total_uncorr_prob}
\end{align}
which recovers the result derived for the ``2-track model'' in \cite{hechenblaikner2021analysis}, where no jitter correlations between tracks were assumed.

\subsection{The linearized analytical model}
In practice there are often moderate but not very high correlations. The limit of small correlation amplitudes  is defined by $\delta_r\ll 1$ and $\epsilon_r\approx 1$ to first order in $\delta_r$.
Linearizing Eq.~\ref{eq:total_corr_prob} for $P_{fail}(\delta_r)$ with respect to $\delta_r$ yields:
\begin{align}
&P_{fail}(\delta_r)\approx\nonumber\\
&\approx\frac{1}{D_t}\int_0^{D_t/2}dx_t\int_{R_d-x_t}^{\infty} dx_n g(x_n)\erfc\left(\frac{R_d-D_t+x_t}{\sqrt{2}\sigma_n}+\delta_r \frac{x_n}{\sqrt{2}\sigma_n}\right)\nonumber\\
&\approx P_{fail,uc}-\frac{\delta_r \sqrt{2}}{\sqrt{\pi}\sigma_n D_t}\int_0^{D_t/2}dx_t e^{-\frac{\left(R_d-D_t+x_t\right)^2}{2\sigma_n^2}} \int_{R_d-x_t}^{\infty} dx_n x_n g(x_n)\nonumber\\
&\approx P_{fail,uc}-\frac{\delta_r}{\pi D_t}\int_0^{D_t/2}dx_t e^{-\frac{\left(R_d-D_t+x_t\right)^2}{2\sigma_n^2}}e^{-\frac{\left(R_d-x_t\right)^2}{2\sigma_n^2}}\nonumber\\
&\approx P_{fail,uc}-\frac{\delta_r\sigma_n}{2\sqrt{\pi} D_t}e^{-\frac{\left(2R_d-D_t\right)^2}{4\sigma_n^2}}\erf\left(\frac{D_t}{2\sigma_n}\right).
\label{eq:total_corr_lin}
\end{align}
Equation~\ref{eq:total_corr_lin} expresses the probability of acquisition failure, linearized with respect to the correlation amplitude $\delta_r$, in terms of basic transcendental functions. As previously done in Eq.~\ref{eq:average_P_fail}, Eq.~\ref{eq:total_corr_lin} can now be averaged over the uncertainty distribution, which amounts to averaging $\delta_r$ because all other parameters in Eq.~\ref{eq:total_corr_lin} do not depend on the radial distance $r$. Using Eqs.~\ref{eq:Rayleigh}, \ref{eq:delta_r} the mean correlation amplitude is obtained:  
\begin{align}
 \delta_{mean}&=\int_0^{\infty} dr\,e^{-\beta r}\frac{r}{\sigma_{uc}^2}e^{-\frac{r^2}{2\sigma_{uc}^2}}\nonumber\\
&=1-e^{\frac{1}{2}\sigma_{uc}^2\beta^2}\sigma_{uc}\beta\sqrt{\frac{\pi}{2}}\erfc\left(\frac{\sigma_{uc}\beta}{\sqrt{2}}\right). \label{eq:delta_mean}
\end{align}
The linearized probability of failure, averaged over the uncertainty distribution, is then found from Eqs.~\ref{eq:average_P_fail},\ref{eq:total_corr_lin},\ref{eq:delta_mean}: 
\begin{align}
P_{fail,av}&=P_{fail,uc}-\frac{\sigma_n}{2\sqrt{\pi} D_t}e^{-\frac{\left(2R_d-D_t\right)^2}{4\sigma_n^2}}\erf\left(\frac{D_t}{2\sigma_n}\right)\times\nonumber\\
&\times\left[1-e^{\frac{1}{2}\sigma_{uc}^2\beta^2}\sigma_{uc}\beta\sqrt{\frac{\pi}{2}}\erfc\left(\frac{\sigma_{uc}\beta}{\sqrt{2}}\right)\right].\label{eq:P_fail_linearized}
\end{align}

\section{Discussion of simulation results}
\label{sec:discussion_of_results}
In this section the derived analytical models are evaluated for the given mission configuration parameters of Table~\ref{tab:simulation_parameters}, unless specified otherwise, and the analytical predictions are compared to the results of Monte Carlo simulations. The observed effects and the validity of previous assumptions are discussed in detail with the aim to give a comprehensive understanding of all correlation effects in this parameter regime.

\subsection{Variation of the scan speed}
\label{sec:Variation_gamma}
\begin{figure}
\centerline{\includegraphics[width=\columnwidth]{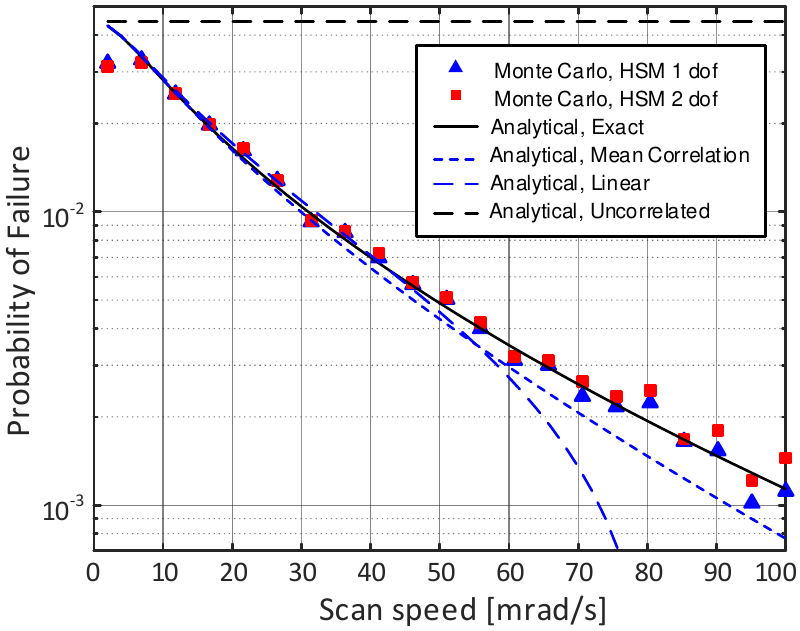}}
\caption  {$P_{fail,av}$ plotted against scan speed $\gamma$ for a track width of $D_t=62.8\,\mu rad$ (black solid line). Results of MC simulations are given by blue triangles (1 jitter dof) and red rectangles (2 jitter dof). The blue-dashed line denotes the linear approximation, the blue-dotted line $P_{fail}(\delta_{mean})$, and the black-dashed line the limit of uncorrelated jitter. MC simulations per data point: $6\times 10^4$.\label{fig_3}}
\end{figure}

Fig.~\ref{fig_3} shows how $P_{fail,av}$ drops with increasing scan speed from an initial value of $\sim 4.5\%$ for scan speeds well below $10\,mrad/s$ to a value of $\sim 0.1\%$ for a scan speed of $100\,mrad/s$. When the scanning beam misses the SC on a certain track due to a large jitter excursion towards an adjacent track, it has a higher probability of hitting the SC on the adjacent track if the excursion persists for at least one spiral revolution (see Fig.~\ref{fig_1}b). Such jitter correlations between tracks increase with scan speed as the mean spiral revolution time becomes increasingly shorter and falls below the jitter correlation time (see Eq.~\ref{eq:correlation_condition}).\\
The black solid line of Fig.~\ref{fig_3} corresponds to the predictions of Eq.~\ref{eq:average_P_fail} and is in remarkably good agreement to the results from Monte Carlo simulations.  The simulations were performed with a Hard-Sphere detection Model (HSM) where the beam is assumed to be detected by the SC if the distance between the beam center and the SC is smaller than the radius of detection $R_d$ at any any time during the spiral scan.  In order to simulate the jitter excursions from the nominal track, the jitter spectrum of Eq.\,\ref{eq:jitter_spectrum} was either applied along radial direction alone (1 dof, blue triangles), or along radial and tangential direction (2 dof, red squares). As also concluded in previous studies \cite{hechenblaikner2022impact,hechenblaikner2023optical}, it is observed that the results of 1 dof and 2 dof are very close, indicating that tangential jitter has no significant impact on $P_{fail}$. This can be intuitively explained by considering that slowly varying tangential displacements will only slightly delay or advance the point in time when the beam passes by the SC, which should have little impact on $P_{fail}$. The assumption that tangential jitter can be neglected was also discussed in section 1\ref{sec:analytical_models_overview} and implicitly considered in the derivation of the ``1d model'' in section 2\ref{sec:model_derivation}, where only beam displacements in radial direction of the uncertainty plane are accounted for.\\
It is observed that in the limit of very low scan speeds the correlated analytical model prediction (solid black line) converges to the ``uncorrelated limit'' (dashed black line) predicted by Eq.~\ref{eq:total_uncorr_prob}. This is expected, because with decreasing scan speed the mean revolution time $T_{mean}$ increases according to Eq.~\ref{eq:T_mean} and becomes larger than the jitter correlation time $\tau_0$, which implies that jitter fluctuations between tracks become uncorrelated. The blue dotted line is computed from the analytical approximation $P_{fail}(\delta_{mean})$ which is obtained by evaluating Eq.~\ref{eq:total_corr_prob} for the mean correlation amplitude $\delta_{mean}$. As can be seen in the figure, this approximation is reasonably close to the exact prediction (black solid line) but does not agree nearly as well with the results of the Monte Carlo simulation. The dashed blue curve is derived from the linearized analytical model which agrees well with the predictions of the exact model for lower scan speeds but starts diverging for larger scan speeds, where correlation amplitudes increase and the linear approximation breaks down.\\
Finally, it is interesting to note that the Monte Carlo simulation results for the lowest scan speed ($\gamma=2\,mrad/s$) are clearly offset from the analytical prediction while the results for larger scan speeds match almost perfectly. This deviation is not a statistical outlier, as one might mistakenly conclude, but the consequence of the fact that for very low scan speeds another regime is entered. It was already mentioned in section 1\ref{sec:analytical_models_overview} that in this regime the jitter fluctuations become uncorrelated on a time-scale shorter than the beam passage time $T_{pass}$, which reduces the failure rates with respect to the predictions of the ``2-track model'' of Eq.~\ref{eq:total_uncorr_prob}. More detailed investigations of this effect, which are outside the scope of this paper, are given in \cite{hechenblaikner2023optical}.

\subsection{Variation of the track-width}
\label{sec:Variation_Dt}
\begin{figure}
\centerline{\includegraphics[width=\columnwidth]{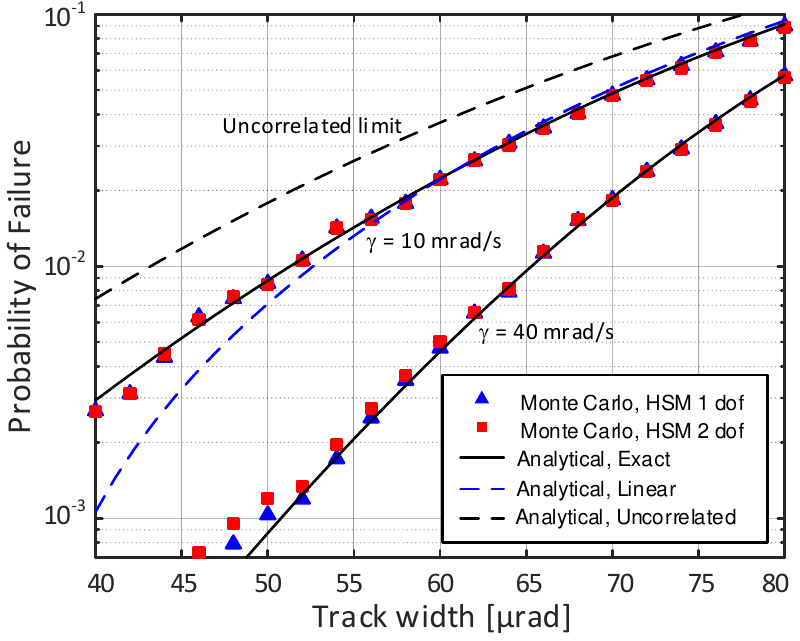}}
\caption  {$P_{fail,av}$ plotted against track width $D_t$ for two different scan speeds of $\gamma=10\,mrad/s$ (upper black line) and $\gamma=40\,mrad/s$ (lower black line). Results of MC simulations are given by blue triangles (1 jitter dof) and red rectangles (2 jitter dof). The blue-dashed line denotes the linear approximation for $\gamma=10\,mrad/s$, the black-dashed line the limit of uncorrelated jitter. MC simulations per data point: $6\times 10^4$ to $10^5$.\label{fig_4}}
\end{figure}

In order to verify the accuracy of the analytical model of Eq.~\ref{eq:average_P_fail}, the variation of $P_{fail,av}$ was studied over a range of track-widths from $D_t=R_d$ to $D_t=2R_d$, which is shown in Fig.~\ref{fig_4}. Remarkably good agreement is found again between the predictions of the correlated analytical model (black solid lines) and the results of Monte Carlo simulations for radial jitter (blue triangles) and radial as well as tangential jitter (red squares). Simulations were made for scan speeds of $\gamma=10\,mrad/s$ and $\gamma=40\,mrad/s$, where the mean correlation amplitudes are given by $\delta_{mean}=0.30$ and $\delta_{mean}=0.71$, respectively. Both curves indicate that the probability of failure increases strongly with increasing track-width, as expected. However, the curve for the lower scan speed ($\gamma=10\,mrad/s$) is closer to the uncorrelated limit predicted by Eq.~\ref{eq:total_uncorr_prob} (black dashed line) because the mean correlation amplitude is smaller. For comparison, the linear approximation predicted by Eq.~\ref{eq:P_fail_linearized} (blue dashed line) is also plotted. It is in good agreement with the exact solution for larger track-widths but diverges noticeably for very low track-widths.

\subsection{Parametric relationships}
\label{sec:parametric_relationships}
It is important to note that the three parameters which define correlation effects ($\sigma_{uc}$, $f_r$, and $\gamma$) only enter the average probability of failure $P_{fail,av}$ through the compound correlation parameter $\eta=\gamma/(\sigma_{uc} f_r)$, i.e. $P_{fail,av}$ is invariant to changes of these parameters as long as the compound parameter $\eta$ remains constant.
This can be be readily proven by considering that the probability of failure depends on correlations through $\delta_r$  which is itself a function of $r f_r/\gamma$ (see Eq.~\ref{eq:delta_r}). Introducing the dimensionless variable $\tilde{r}=r/\sigma_{uc}$ in the integral of Eq.~\ref{eq:average_P_fail} yields
\begin{equation}
P_{fail,av}=\int_0^{\infty} d\tilde{r} P_{fail}\left(\tilde{r} \sigma_{uc}f_r/\gamma\right)\tilde{r}e^{-\frac{\tilde{r}^2}{2}}=\int_0^{\infty} d\tilde{r} P_{fail}\left(\tilde{r}/ \eta\right)\tilde{r}e^{-\frac{\tilde{r}^2}{2}},
\label{eq:P_fail_invariant}
\end{equation}
which explicitly shows the dependency of $P_{fail,av}$ on the compound parameter $\eta$.
As an example, if the width of the uncertainty distribution $\sigma_{uc}$ is reduced by a certain factor, the scan speed $\gamma$ can also be reduced by the same factor without changing the probability of failure.

\begin{figure}
\centerline{\includegraphics[width=\columnwidth]{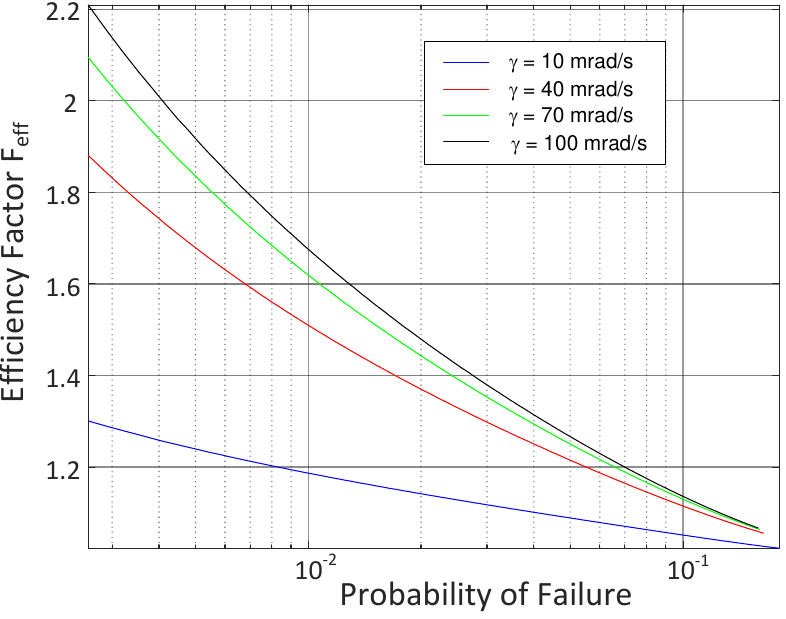}}
\caption  {Efficiency factor $F_{eff}$ plotted against probability of failure for 4 different scan speeds with $\gamma=10,40,70,100\,mrad/s$ (bottom to top), respectively.\label{fig_5}}
\end{figure}

Now assume that for a certain track width $D_t'$ a probability of failure $P_{fail,uc}(D_t')$ is obtained if there are no correlations. The track-width $D_t$ is then determined where the same probability of failure $P_{fail,av}(D_t)$ is obtained in case of jitter correlations. The ``correlation efficiency factor'' $F_{eff}=D_t/D_t'$ specifies a metric for the performance improvement obtained through the effect of correlations:
\begin{align}
P_{fail,av}(D_t)&=P_{fail,uc}(D_t')\nonumber\\
F_{eff}&=D_t/D_t'
\end{align}
As correlations tend to decrease the probability of failure, the track-width can generally be increased compared to the uncorrelated case, which reduces the mean acquisition time in inverse proportion to the track-width. Therefore, the performance is expected to improve and $F_{eff}$ is expected to increase with the correlation parameter $\eta$. This is what is observed in Fig.~\ref{fig_5} which plots $F_{eff}$ against probability of failure for 4 different scan speeds.
For example, in order to obtain a failure probability of 1\%, the track width is increased by a factor $F_{eff}=1.51$ for $\gamma=40\,mrad/s$ and by $F_{eff}=1.67$ for $\gamma=100\,mrad/s$ with respect to the case that no correlations are present. 

\subsection{Mean search time for multiple scans}
\label{sec:mean_acquisition_time}
\begin{figure}
\centerline{\includegraphics[width=\columnwidth]{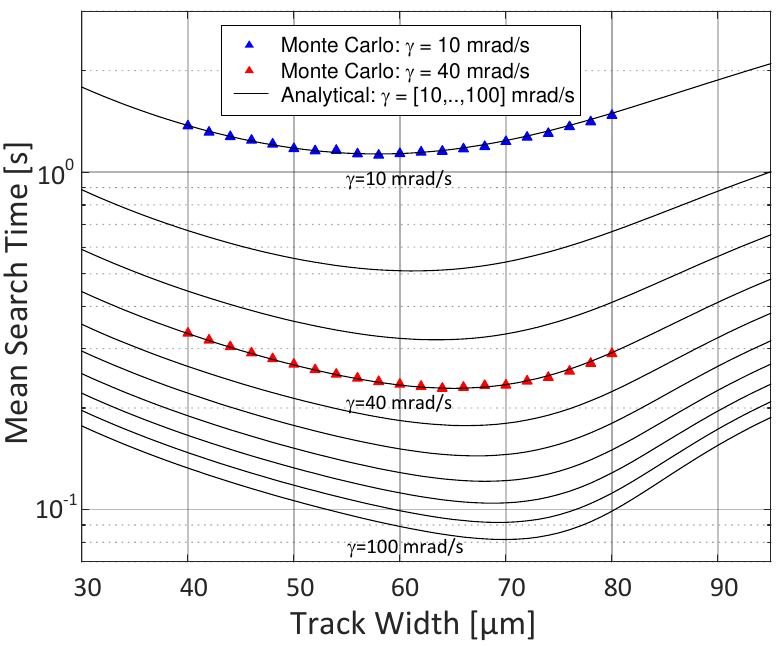}}
\caption{Mean search time for multiple scans $T_{ms}$ of Eq.~\ref{eq:mean_acquisition_time} plotted against track-width $D_t$ for various scan speeds from $10\,mrad/s$ to $100\, mrad/s$ (black lines). Results from MC simulations are given by the blue triangles ($\gamma=10\,mrad/s$) and red triangles ($\gamma=40\,mrad/s$) for $6\times 10^4$ and $10^5$ simulations per data point, respectively.\label{fig_6}}
\end{figure}

It is generally possible to repeat the spiral scan in case it fails the first time. Designing the acquisition search for a low probability of failure $P_{fail}$ (of a single scan) in presence of jitter requires choosing a narrow track width $D_t$, which increases the mean search time for a single successful scan given by $T_{1s}=2\pi\sigma_{uc}^2/(D_t\gamma)$ \cite{hechenblaikner2021analysis}. Choosing a larger track-width increases $P_{fail}$, which may require performing another scan if the beam has not been detected during the entire scan across the uncertainty region of radius $R_{uc}$. This also increases the mean search time for multiple scans $T_{ms}$ which accounts for the total time required if a search scan must be repeated due to failure. Therefore, for best performance, the right balance between two counter-acting effects must be found: if the track-width increases (or, alternatively, the beam overlap decreases), $T_{1s}$ decreases on the one hand, but $P_{fail}$ increases on the other hand. The authors of \cite{li2011analytical} derived an expression for $T_{ms}$. After some re-arrangements it is found to essentially depend on $T_{1s}$ and $P_{fail}$ as follows \cite{hechenblaikner2023optical}:
\begin{equation}
T_{ms}=T_{1s}\left(1+\frac{F_{uc}^2}{2}\frac{P_{fail}}{1-P_{fail}}\right),
\label{eq:mean_acquisition_time}
\end{equation}
where the factor $F_{uc}=R_{uc}/\sigma_{uc}$ relates the radius of the scan region to the width of the uncertainty distribution. It is typically chosen as $F_{uc}=3.44$, yielding a 99.73~\% probability that SC2 is located within the radius $R_{uc}$ according to Eq.~\ref{eq:Rayleigh}. Using the values of $P_{fail,av}$ predicted from Eq.~\ref{eq:average_P_fail} and substituting them into Eq.~\ref{eq:mean_acquisition_time}, the mean search time for multiple scans $T_{ms}$ is obtained. The latter is plotted against the track-width for various scan speeds from $10\,mrad/s$ to $100\,mrad/s$, given by the black solid lines in Fig.~\ref{fig_6}. The results of Monte Carlo simulations (blue and red triangles) are found to be in very good agreement to the analytical predictions.\\
From these curves it is easy to determine the minimum search time $T_{ms,min}$ and the associated track-width $D_{t.min}$ which optimize the acquisition architecture for a given scan speed. These parameters are found to change from $T_{ms,min}=1.13\,s$, $D_{t,min}=58\,\mu rad$ (overlap of $0.55\,R_d$) for $\gamma=10\,mrad/s$ to $T_{ms,min}=0.08\,s$, $D_{t,min}=70\,\mu rad$ (overlap of $0.25\,R_d$) for $\gamma=100\,mrad/s$. This indicates that the optimal beam overlap $OV$ decreases with increasing scan speed to smaller and smaller values, owing to the emergence of correlations that mitigate the impact of beam jitter.

\section{Conclusion}
In this paper a comprehensive overview is given of important aspects to be considered in the development of a probabilistic model for the spatial acquisition of optical links.
An analytical model for the probability of acquisition failure ($P_{fail}$) under the influence of beam jitter is derived, where jitter correlation effects that depend on the spectral shape $S(f)$, the scan speed $\gamma$, and the width of the uncertainty distribution $\sigma_{uc}$ are considered. 
Correlations of the jitter amplitude between two successive tracks of the search spiral can greatly reduce $P_{fail}$. They emerge if the width of the jitter auto-correlation function is larger than the mean spiral revolution time $\tau_0>T_{mean}$\cite{hechenblaikner2022impact}, i.e. if the jitter band-width is sufficiently small (increasing $\tau_0$) and the scan speed sufficiently fast (decreasing $T_{mean}$), which is a parameter regime accessible to typical optical communication missions in space.\\
In the derivation the jitter fluctuations were decomposed into correlated (parallel) and uncorrelated (normal) components, whose amplitudes could be obtained from the normalized jitter auto-correlation function evaluated for the spiral revolution time. This allowed obtaining closed-form analytical solutions. In the limit of fully correlated jitter $P_{fail}$ was found to approach $0$, and in the limit of uncorrelated jitter the previous result of a simpler model \cite{hechenblaikner2021analysis} was obtained. Linearizing the analytical model with respect to the correlation amplitude enabled $P_{fail}$ to be expressed in terms of basic transcendental functions.\\
Very good agreement was found between the predictions of the analytical model and the results of Monte Carlo simulations, which verified the validity of the analytical model in the narrower sense and the conceptual understanding of correlation effects in the broader sense. Model predictions confirmed that increasing the scan speed leads to a decrease of $P_{fail}$ from the limit of uncorrelated jitter towards ever smaller values, as expected. For correlated jitter a similar dependency of $P_{fail}$ on the track-width $D_t$ was found as observed for uncorrelated jitter, albeit with a shift towards lower probabilities depending on the correlation strength.\\
It was shown that correlation parameters only enter the probability of failure through a compound scaling parameter $\eta=\gamma/(\sigma_{uc}f_r)$ and the correlation efficiency factor $F_{eff}$ was introduced which defines the performance increase relative to the uncorrelated jitter limit. Finally, the analytical model allowed calculating the mean search time for repetitive scans and determining the minimum time associated with the optimal choice of track-width for a given scan speed.\\
In conclusion, the analytical model presented in this paper accounts for jitter correlation effects, while previous analytical models and most simulations only consider the total jitter power, which may lead to large errors in the predicted probability of acquisition failure for certain mission configurations. Therefore, this model may help improve and optimize the acquisition architecture in performance critical applications, in particular those employing a fast scanning terminal accommodated on a platform defined by a low-bandwidth jitter spectrum.

\begin{backmatter}
\bmsection{Acknowledgments} The author gratefully acknowledges fruitful discussions with R\"udiger Gerndt (Airbus).

\bmsection{Disclosures} The author declares no conflicts of interest.

\bmsection{Data availability} The data underlying the results presented in this paper were generated using analytical formulas presented in the text as well as Monte Carlo simulations. Simulation data can be provided upon reasonable request.

\end{backmatter}


%
%
\end{document}